\documentclass[aps,pra,preprint,superscriptaddress]{revtex4-1}
\usepackage{graphicx}
\usepackage{xcolor}
\usepackage{amsmath}
\usepackage{hyperref}
\usepackage{dcolumn}
\usepackage{bm}
\begin{document}
\title{Simulating the structural diversity of carbon clusters across the planar to fullerene transition}
\author{Ma\"elle A. Bonnin}
\affiliation{Institut des Sciences Mol\'eculaires d'Orsay, Univ. Paris-Sud, Universit\'e Paris-Saclay, 91405 Orsay, France.}
\author{Cyril Falvo}
\affiliation{Institut des Sciences Mol\'eculaires d'Orsay, Univ. Paris-Sud, Universit\'e Paris-Saclay, 91405 Orsay, France.}
\affiliation{Univ. Grenoble Alpes, CNRS, LIPhy, 38000 Grenoble, France}
\author{Florent Calvo}
\affiliation{Univ. Grenoble Alpes, CNRS, LIPhy, 38000 Grenoble, France}
\author{Thomas Pino}
\affiliation{Institut des Sciences Mol\'eculaires d'Orsay, Univ. Paris-Sud, Universit\'e Paris-Saclay, 91405 Orsay, France.}
\author{Pascal Parneix}
\email{pascal.parneix@u-psud.fr}
\affiliation{Institut des Sciences Mol\'eculaires d'Orsay, Univ. Paris-Sud, Universit\'e Paris-Saclay, 91405 Orsay, France.}
%
%
\begin{abstract}
Together with the second generation REBO reactive potential, replica-exchange molecular dynamics simulations coupled with systematic quenching were used to generate a broad set of isomers for neutral C$_n$ clusters with $n=24$, 42, and 60. All the minima were sorted in energy and  analyzed using order parameters to monitor the evolution of their structural and chemical properties. The structural diversity measured by the fluctuations in these various indicators is found to increase significantly with energy, the number of carbon rings, especially 6-membered, exhibiting a monotonic decrease in favor of low-coordinated chains and branched structures. A systematic statistical analysis between the various parameters indicates that energetic stability is mainly driven by the amount of sp$^2$ hybridization, more than any geometrical parameter. The astrophysical relevance of these results is discussed in the light of the recent detection of C$_{60}$ and C$_{60}^+$ fullerenes in the interstellar medium.
\end{abstract}
\maketitle

\section{Introduction}
From a few atoms to bulk matter, carbon clusters show a significant ability to hybridize in sp, sp$^2$ or sp$^3$ chemical bonds, reflecting at finite size the wide allotropy of bulk carbon matter. Depending on experimental conditions, carbon clusters can be produced into a very large variety of isomers that have been probed by many groups for more than two decades~\cite{Helden1991,Helden1993,lopez97,Orden1998,HANDSCHUH1995,Lifshitz2000,Kent2000,Ueno2008,Kosimov:2010aa,straatsma17,Kono18}. Below about 20 carbon atoms, (1D) chains and (2D) rings have been identified as the most stable isomers~\cite{Helden1991,Orden1998,Brito18} while (3D) fullerenes were shown to be the most stable form of larger carbon clusters~\cite{Tomanek91,Kent2000}.

Recently there was a surge of interest in the relaxation dynamics in carbon-based clusters following their brief excitation typically produced by energetic particle collisions \cite{Delaunay:2015aa,Simon18}
or short laser pulses \cite{marciniak15,ji17,Zhen18}. Such experiments are largely motivated by the increasing evidence from astronomical observations that pure and hydrogenated carbon clusters are indeed present in the interstellar medium (ISM). While only the smallest molecules up to C$_5$ were conclusively detected until 2001 \cite{Bernath:1989aa,Maier2001}, the much more organized fullerenes C$_{60}$ and probably C$_{70}$ were observed in the ISM owing to their very characteristic vibrational infrared emission bands~\cite{Sellgren2010,Cami2010}. The possible detection of the cation C$_{60}^+$ from its infrared emission bands was also suggested~\cite{Berne2013}. These fullerene bands accompany the so-called aromatic infrared bands (AIBs), which trace polycyclic aromatic aliphatic mixed hydrocarbons widely observed in the ISM \cite{LEGER1984,Allamandola1985,Sloan2007,Pino2008,Acke2010,Sloan2014,Pilleri2015}.
Interestingly, infrared spectroscopy was also used to characterize smaller carbon clusters in laboratory experiments by Straatsma and coworkers \cite{straatsma17}.

The high level of chemical organization of fullerenes necessarily raises questions regarding their formation under the harsh conditions of astrophysical environments. So far, essentially two scenarios have been proposed to explain the presence of such molecules in the ISM. The so-called bottom-up hypothesis, in which fullerenes would be formed by coalescence of smaller entities~\cite{Yamaguchi:1998aa,Irle:2003aa,Zheng:2005aa}, is rather unlikely in the ISM owing to its extremely low densities of matter \cite{Berne2015}. In the top-down scenario, fullerenes originate from the decay of larger compounds subject to energetic excitation (cosmic rays, shocks, VUV irradiation...) followed by stepwise isomerization \cite{Berne2015,Chuvilin10,Montillaud13,Pietrucci14,Zhang13}. In this respect, amorphous carbon clusters were suggested to play a possibly important role \cite{Powles:2009aa,Jones2017,sinitsa17}. Such formation pathways thus question the existence of products that are concomitantly formed and may be stable in space, spanning the range between amorphous and fullerenic carbon clusters.

Experimental observation of the possible formation mechanisms will necessarily rely on spectroscopy, and theoretical interpretation of the measured spectra requires appropriate candidate structures. Unfortunately, the typical approach until now has usually been biased towards certain chemical types such as polyaromatic compounds \cite{Bauschlicher:2018aa}, only limited effort being devoted to arbitrary or amorphous conformations despite their known astrophysical relevance \cite{duley12,Jones2013,Jones2017,Bouteraon2019}.
According to the theoretical study by Kent and coworkers \cite{Kent2000}, 24 carbon atoms are needed to produce the first polyaromatic flake, and stable fullerenes arise above the size 26, although not necessarily the most stable conformation. Size 60 is notorious to support fullerenes as their most stable structure. In the absence of known magic numbers in this size range, and as common in clusters physics, 2D and 3D structures are thus likely to coexist \cite{Helden1991}.

The present article aims at exploring the structural
diversity of carbon clusters as a function of their most important feature, namely their internal energy, in the size range where they undergo the flake-to-fullerene transition. Although the degree of chemical ordering is obviously expected to vary with increasing excitation energy, the extent of chemical and conformational variety remains undocumented so far, and as far as we are aware our fundamental study represents the first attempt to construct a library of carbon cluster structures in an unbiased way, not focusing on the lowest-energy structures only but giving as much attention to higher-energy isomers that could be formed on the interstellar fullerenic road .

To reach this goal we have systematically explored the conformational landscapes of selected carbon clusters C$_n$ across the planar to fullerene transition and containing $n=24$, 42, or 60 atoms, using advanced molecular dynamics methods and systematic quenching. 
The numerous isomers thus obtained were sorted and analyzed using a range of structural order parameters, some of them to quantify the nature of the chemical bonds within the cluster. Because this computational investigation is highly statistical, we relied on a simplified but realistic description of the potential energy surface based on the second generation reactive REBO bond-order potential \cite{Brenner:2002xy}. Such an approach has already been used in the past to study the structure of carbon clusters in the size range up to 55 atoms by Kosimov and coworkers \cite{Kosimov:2010aa}, who predicted a transition from ring structures to graphene flakes occuring above 18 atoms, without reporting any fullerene. In contrast with earlier computational investigations, our approach here is highly statistical and does not focus on the lowest-energy structures.
The results thus obtained shed light on the local chemical ordering and global structural arrangement of carbon clusters as their energy varies down to low-energy polyaromatic isomers.\par

The article is organized as follows. In the next section, we describe the different computational tools used for generating and characterizing the structural diversity of selected carbon clusters at finite internal energy. The
results are presented and discussed in Section III in relevance to astrophysical implications. In Section IV we present a systematic statistical analysis in which we correlate the relative energetic stability of the various conformers to their structural and chemical features. Some concluding remarks are finally given in Section V together with ongoing or future extensions.
\par

\section{Methods}

\subsection{Sampling the conformational landscape}

The energy landscape of carbon clusters C$_n$ is characterized by an exponentially increasing number of local minima and transition states with increasing size $n$, and cannot be sampled exhaustively as soon as this number exceeds a few tens. Furthermore, the barriers separating the various local minima or even more distant funnels on the landscape are likely to be high and make traditional simulation methods based on molecular dynamics or Monte Carlo methods poorly efficient. More importantly, we are not interested in addressing the global optimization problem specifically by focusing on the lowest-energy structures, but rather to explore conformations that are physically relevant also at high energies, as could be produced e.g. upon photonic or collisional excitation, though still under isolated conditions.

We used replica-exchange molecular dynamics (REMD) simulations \cite{Swendsen:1986aa,Marinari:1992aa} to circumvent this broken ergodicity issue and achieve a broader sampling of the potential energy surface over extended energy ranges.

REMD simulations were performed using the LAMMPS program~\cite{Plimpton:1995aa} by propagating $M$ trajectories at fixed temperatures $T_i$ ($i=1\dots M$) and occasionally attempting exchanges between the two configurations ${\bf R}_i$ and ${\bf R}_j$ of neighboring replicas $i$ and $j=i\pm 1$. Such an exchange is accepted using the following acceptance rule~\cite{Sugita:1999aa}
\begin{equation} \label{exrates}
\mbox{acc}({\bf R}_i\rightleftharpoons {\bf R}_j) =\text{min}\left \{ 1,\exp \left[(\beta _i-\beta _j)(E(\mathbf{R}_i)-E(\mathbf{R}_j))\right] \right \}
\end{equation}
where we have denoted $\beta _i=1/(k_BT_i)$ and $E({\bf R})$ the potential energy at configuration ${\bf R}$, discussed below.

The efficiency of this exchange process depends on the overlap between the thermal distributions at temperatures $T_i$ and $T_j$, which itself is driven by various factors, primarily the two temperatures but also the size $n$ which affects the width of the individual distributions. It is thus important to carefully choose the set of temperatures $\{ T_i \}$ for each cluster size, taking also into account the need for the upper temperature $T_M$ to be high enough to ensure an efficient exploration of configurational space, still
below the vaporization temperature. Since the thermal distributions both shift and broaden with increasing temperature, the difference between successive temperatures $T_{i+1}-T_i$ must also increase with $i$. Here we employed a geometrical progression in $T_i$, namely $T_{i+1}=\alpha T_i$, which is optimal in the harmonic limit~\cite{Mitsutake:2001aa}, to which we have added manually some temperatures to increase the exchange probability. 

The MD trajectories were integrated with a $0.1$ fs time step. A Nos\'e-Hoover thermostat was used to keep temperatures constant with a damping constant of $10$~fs. Exchange between nearest replicas were attempted once every $2\times 10^4$~MD steps and the simulation was propagated for 100~ns. Configurations were periodically saved for further structural analysis and quenching every 4~ps, resulting in a total of 25\,000 structures per replica.

During the simulations it was also important to forbid fragmentation as it is an irreversible process preventing the correct sampling of size-selected compounds. Here we used a simple spherical harmonic potential container with a 500 eV$/$\AA$^{2}$ spring constant and a radius $R_s$ whose value was chosen differently for the three cluster sizes $n$ to reach a common density of 1.7~g/cm$^{3}$, relevant for disordered polyaromatic materials such as soot~\cite{Ouf2015}. The initial conditions of the REMD simulations were taken as the global energy minima for each cluster, also denoted as reference structures.

\subsection{Potential energy surface for carbon clusters}

The systematic production of large samples of minima at finite temperature, in which chemical bonds are easily broken and reformed, requires an efficient but chemically realistic method for the potential energy surface $E({\bf R})$. Here methods with an explicit treatment of electronic structure or based on first principles are not practical, especially considering the possible multireference character of small carbon clusters that would make the solution to the electronic problem already cumbersome~\cite{Montagnon:2007aa}.

Only an explicit potential energy surface is currently able to handle the tremendous number of configurations gathered with the currently available computational resources, approximate schemes based on tight-binding~\cite{Wang:1995bh,Van-Oanh:2002oq} or density-based tight-binding~\cite{Rapacioli2015} remaining still too expensive for the present rather large clusters. A few realistic potentials are available for carbon, which correctly account for bond breaking and formation and the various hybridization environments displayed by carbon \cite{Tersoff:1988aa,Brenner:1990ty,Brenner:2002xy,Stuart:2000fk,Marks:2000aa,Ghiringhelli:2005fk,Zhou:2015ab}. Here we have chosen the adaptive the second-generation reactive bond order (REBO) potential of Brenner \cite{Brenner:2002xy}. Brenner type potentials have been notably used to study energetic and mechanical properties of nanotubes \cite{Ni:2002aa}, the formation process of fullerenes~\cite{Yamaguchi:1998aa,Makino:1997aa} but also to describe reactivity and formation of pure carbon clusters or hydrocarbons in the astrophysical context ~\cite{Patra:2014aa,Delaunay:2015aa}, making it a natural choice also for the present investigation. Moreover it has been shown that the REBO potential gives better results to describe small carbon clusters compared with other bond-order potentials~\cite{Lai:2016aa}.

\subsection{Reference structures}

Our REMD exploration requires some initial structures. We have thus conducted a distinct search for the lowest-energy configurations of C$_{24}$ by comparing various remarkable structures such as polycyclic, chains and rings conformers. In agreement with the earlier study by Kent {\em et al.} \cite{Kent2000}, the lowest-energy structure found for C$_{24}$ is the planar, fully dehydrogenated coronene with $D_{6h}$ point group. For C$_{60}$ the natural reference structure is buckminsterfullerene with point group $I_h$, besides a set of 1811 alternative but higher energy fullerenes~\cite{Sure:2017aa}. 
Finally, C$_{42}$ was chosen as an intermediate size cluster with a propensity to form fullerenes. For this cluster, 45 nonequivalent fullerene isomers could be identified~\cite{Fowler} and the REBO potential predicts that the most stable isomer is the only fullerene with $D_3$ symmetry, in accordance with DFT calculations~\cite{Sun:2005aa,Malolepsza:2009aa} but at variance with the earlier study by Kosimov {\em et al.} \cite{Kosimov:2010aa} who found a graphene flake as for C$_{24}$.

These reference structures are depicted in Figs.~\ref{fig:structures}-C$_{24}$-a), ~\ref{fig:structures}-C$_{42}$-a), and~\ref{fig:structures}-C$_{60}$-a) and their absolute binding energies obtained with REBO are given in Table~\ref{tab:energiesref}.
\begin{table}
\begin{tabular*}{\linewidth}{@{\extracolsep{\fill}} ccc}
\hline
\hline
Reference structure & Point group & Binding energy (eV/atom) \\
\hline
C$_{24}$ & $D_{6h}$ & 6.237 \\
C$_{42}$ & $D_3$ & 6.614 \\
C$_{60}$ & $I_h$ & 6.842 \\
\hline
\hline
\end{tabular*}
\caption{Point group and absolute binding energy of the reference structures of C$_{24}$, C$_{42}$ and C$_{60}$ obtained with the REBO potential.}
\label{tab:energiesref}
\end{table}
For buckminsterfullerene, the binding energy is reasonably close to the experimentally known value of 6.7~eV/atom \cite{dresselhaus}.

\subsection{Computational details}

The number of temperatures allocated for the REMD trajectories, their lowest and highest values and the container radius employed in the simulations are given for each system size in Table \ref{MT}.
\begin{table}
\begin{tabular*}{\linewidth}{@{\extracolsep{\fill}} ccccc}
\hline
\hline
Cluster & $T_{{\rm{min}}}$ (K) & $T_{{\rm{max}}}$ (K) & $M$ & $R_{s}$ (\AA) \\
\hline
C$_{24}$ & 1500 & 6500 & 12 & 4.0707 \\
C$_{42}$& 2500 & 6500 &  14 & 4.9030 \\
C$_{60}$ & 2500 & 6500 & 16 & 5.5209 \\
\hline
\hline
\end{tabular*}
\caption{\label{MT} Parameters used for determination of temperatures in the REMD simulations.}
\end{table}
For all cluster sizes and for each of the $M$ replicas, 25\,000 configurations were generated. Given the numbers of replicas employed for each system, a total of 300\,000 instantaneous configurations were kept for further analysis for C$_{24}$, 350\,000 instantaneous configurations for C$_{42}$ and 400\,000 configurations for C$_{60}$.

\subsection{Identification of stable structures}

From the large sets of instantaneous configurations gathered at finite temperature, the local minima were obtained by systematic quenching using here the Hessian-free truncated Newton method as implemented in LAMMPS~\cite{Plimpton:1995aa} and disregarding the hard-wall spherical container. Only connected structures were subsequently kept for further analysis, disconnected configurations being discarded. Here fragments are identified using a maximum nearest neighbor distance of 1.85~\AA. In order to speed up the analysis, two locally minimized structures were further considered to be identical if their energy difference lies below 0.01~meV, in which case the highest was discarded as well. After this optimization and screening stage, the numbers of distinct configurations saved for further processing was equal to 51\,901, 240\,305, and 236\,394 for C$_{24}$, C$_{42}$, and C$_{60}$, respectively. We interpret this slightly smaller number for the larger cluster size as reflecting its more magic character, the diversity of conformers at a same upper temperature (here 6500~K) being lower than in a non-magic system such as C$_{42}$~\cite{Walesbook}. However, the energy distributions shown below are essentially robust when considering only the first or second half of the configurations harvested during the REMD trajectories, suggesting that our samples are statistically representative.
Yet, the smaller number of distinct configurations for C$_{60}$ would probably require further scrutiny, and it could be useful to apply alternative approaches employing systematic local minimization such as basin-sampling \cite{bogdan06} in future extensions of this work dealing with even more complex systems.

\subsection{Structural analysis}
\label{sec:struct}

Both local and global parameters were used to characterize the various isomers obtained for the three clusters sizes. Global parameters provide information about the overall shape and atomic distribution around the center of mass, while local parameters give insight into the chemical arrangement at the atomic level.

For a $n$-atom cluster C$_n$ with equilibrium configuration ${\bf R}=\{ {\bf r}_i \}$ for $i=1\dots n$, the $3\times 3$ gyration tensor ${\mathbf{Q}}$ is defined from its components $Q^{\alpha\beta}$ as \cite{Solc:1971aa,Theodorou:1985aa,Blavatska:2010aa,Calvo:2012aa}
\begin{equation} \label{girtens}
Q^{\alpha\beta}=\frac{1}{n}\sum_{i=1}^{n}( r_i^\alpha-\bar r^\alpha)(r_j^\beta-\bar r^\beta), \quad \alpha,\beta=1,2,3,
\end{equation}
where $r_i^\alpha$ are the Cartesian coordinates ($x$, $y$ or $z$) of atom $i$, $\bar r^\alpha$ are the corresponding coordinates for the cluster center of mass \cite{Blavatska:2010aa}. Three rotationally invariant quantities can be defined from the tensor ${\bf Q}$ that respectively measure the geometrical extension, the asphericity, and the prolateness of the atomic distribution. The squared radius of gyration $R_g^2$ is first defined by
\begin{equation}
\label{R2}
R_g^2=\frac{1}{n}\sum_{i=1}^{n}({\bf r}_i - {\bf \bar r})^2  =\text{Tr}\,\mathbf{Q},
\end{equation}
where $\text{Tr}\,\mathbf{Q}$ stands for the trace of the gyration tensor. The other two quantities are defined from the traceless tensor $\mathbf{D}=\mathbf{Q} -\frac{1}{3}\mathbf{I}\text{Tr} \mathbf{Q}$, where $\mathbf{I}$ is the 3$\times$3 identity matrix. The asphericity parameter $A_3$ is defined by \cite{Calvo:2012aa} 
\begin{equation}  
\label{A3}
A_3=\frac{3}{2}\frac{\text{Tr}\,\mathbf{D}^2}{(\text{Tr}\,\mathbf{Q})^2},
\end{equation}
The asphericity varies from 0 for a purely spherical system to 1 for a perfectly linear structure. Finally, the prolateness $S$ is given for  configuration $\mathbf{R}$ by 
\begin{equation}
\label{S}
S = 27 \, \frac{\text{det}\,\mathbf{D}}{(\text{Tr}\,\mathbf{Q})^3},
\end{equation}
and varies from $S=-1/4$ for a perfectly planar disk (\textit{oblate} structure) to $S=2$ for a perfectly linear chain (\textit{prolate} structure).

The tendency to form hollow, planar or close-packed structures was investigated also from the radial density $\rho(r)$ around the center of mass,
\begin{equation}
\rho(r>0) = \frac{1}{4\pi r^2} \sum_{i=1}^{n} \delta(r-\left|{\bf r}_i-\bar{\bf r}\right|),
\end{equation}
where $\delta$ refers to the Dirac function.

Another useful information for pure carbon clusters having a tendency for sp and sp$^2$ hybridization is the number of rings of each size. Here we counted the numbers of 3-, 5-, 6-, and 7-membered rings from the analysis of nearest neighbor atoms connectivity and denote them as R$_\ell$ where $\ell$ is the length of the ring.

Turning now to local order parameters, the hybridization state of each atom was quantified for all stable configurations ${\bf R}$ of the databases. In many computational studies that do not explicitly compute the electronic structure~\cite{Galli:1989aa,Galli:1990aa,Wang:1993vn,Bhattarai:2017fk,Dozhdikov:2017zr} hybridization is defined based on coordination numbers only. However, such a definition cannot account for the chemical complexity and possibly reactive atoms that are under coordinated. Here we use both coordination and geometric information to assign hybridization states. More precisely, we define $N_i$ the coordination number of atom $i$, neighbors lying within 1.85~\AA\ from atom $i$, and we evaluate all angles in which atom $i$ is a vertex. There are $M_i=N_i(N_i-1)/2$ such angles which we denote by $\theta_k$ for $k=1\dots M_i$, leaving the dependence on atom $i$ as implicit. Atom $i$ is then said to be in sp hybridization state provided that it is not overcoordinated and the angles in which it is central are close enough to 180$^\circ$:
\begin{equation}
\text{sp} \text{: }
\begin{cases}
N_i = 1 \text{ or } 2, \\
\theta_k > 170^\circ \quad \forall\ k.
\end{cases}
\end{equation}
Likewise, the atom is assigned sp$^2$ hybridization with the following criteria
\begin{equation}
\text{sp}^2 \text{: }
\begin{cases}
N_i = 2\text{ or } 3,\\
95^\circ < \theta_k < 135^\circ \quad \forall\ k,\\
\text{Var}(\theta_k) < 20^\circ,
\end{cases}
\end{equation}
where $\text{Var}(\theta_k)$ denotes the variance of all angles for which $i$ is a vertex. Here again an upper limit for the angular boundary of 20$^\circ$ is chosen in such a way as to include atoms involved in hexagonal and pentagonal rings as properly sp$^2$. The variance limit ensures that for $N_i=3$ all 4 atoms remain close to a common plane. It was notably chosen to include fullerene structures (for buckminsterfullerene 
Var$(\theta_k)=6^\circ$). 
Finally atom $i$ is considered to be in a sp$^3$ hybridization state accordingly with
\begin{equation}
\text{sp}^3 \text{: }
\begin{cases}
N_i = 4, \\
85^\circ < \theta_k < 125^\circ \quad \forall\ k,\\
\text{Var}(\theta_k) < 20^\circ.
\end{cases}
\end{equation}
Here the variance condition on the angles is required to ensure the atom has a 3D tetrahedral environment. With the above definitions, it may occur that a given atom is neither sp, sp$^2$ or sp$^3$, in which case it will  be referred to as ambiguous. This notably occurs for ringlike structures containing too few atoms, hence producing angles that exceed 10$^\circ$, one such example being illustrated in Fig.~\ref{fig:structures}-C$_{24}$-d). With the present criterion on angles, at least 36 atoms would be needed in the perfect ring for the sp$^2$ assignment to be recovered.

Finally, as our last local structural quantity we have determined the pair correlation function $g(r)$ from
\begin{equation}
g(r) = \frac{1}{4\pi r^2 \rho_S} \sum_{i<j} \delta(r - \left| {\bf r}_i-{\bf r}_j \right|),
\end{equation}
where $\rho_S=3n/4\pi R_S^3$ is the original density used to constrain the REMD simulations. The pair correlation function ignores the overall distribution of atoms within the container and thus provides a complementary information to the radial distribution. Furthermore it is indirectly sensitive to the chemical nature of the bonds, the various hybridization states being related to different equilibrium distances in carbon-carbon bonds. For example, with the REBO potential the CC distance of acetylene (C$_2$H$_2$) is 1.21~\AA, increasing to 1.32~\AA\ in ethylene (C$_2$H$_4$) and to 1.54~\AA\ in ethane (C$_2$H$_6$). These values are in very satisfactory agreement with quantum chemical calculations at the DFT/M06-2X/6-311++G** level, which for the same molecules yield 1.1981, 1.3260, and 1.5254~\AA, respectively.

\section{Results and discussions}

\begin{figure}
\centering
\includegraphics{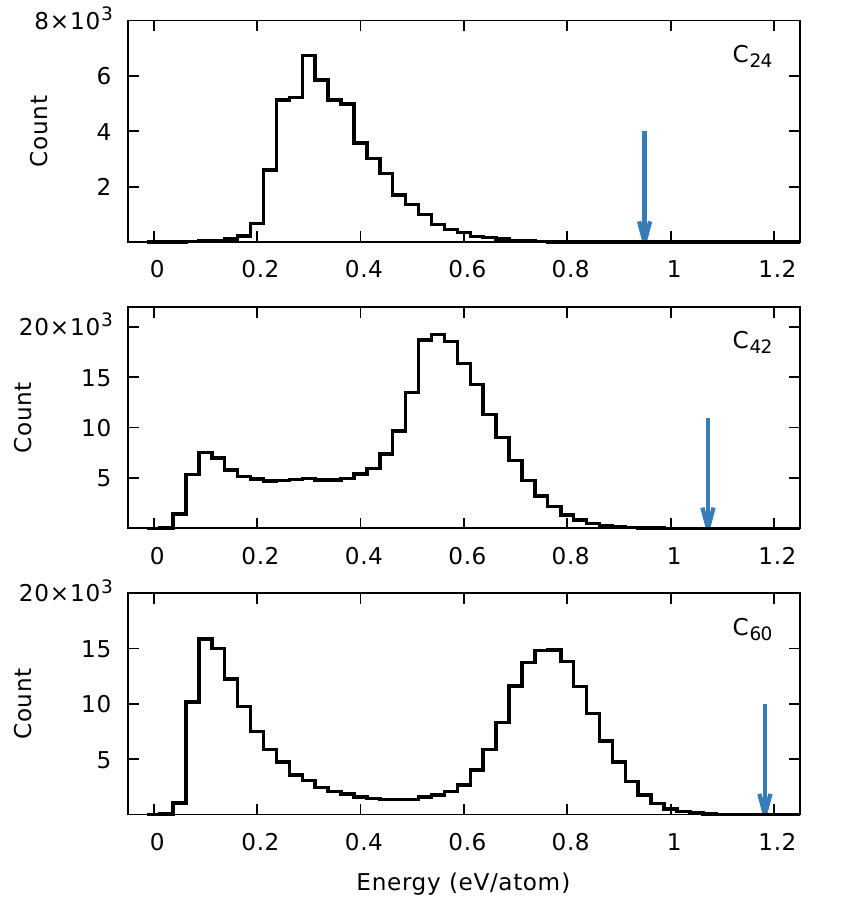}
\caption{(Color online) Energy distributions of quenched structures for C$_{24}$, C$_{42}$, and C$_{60}$ relative to the corresponding reference structure (lowest isomer). The vertical blue arrows locate the highest energy isomers found in the REMD simulations.}
\label{fig:distribEp_min}
\end{figure}

\begin{figure}
\centering
\includegraphics[width=8.5cm]{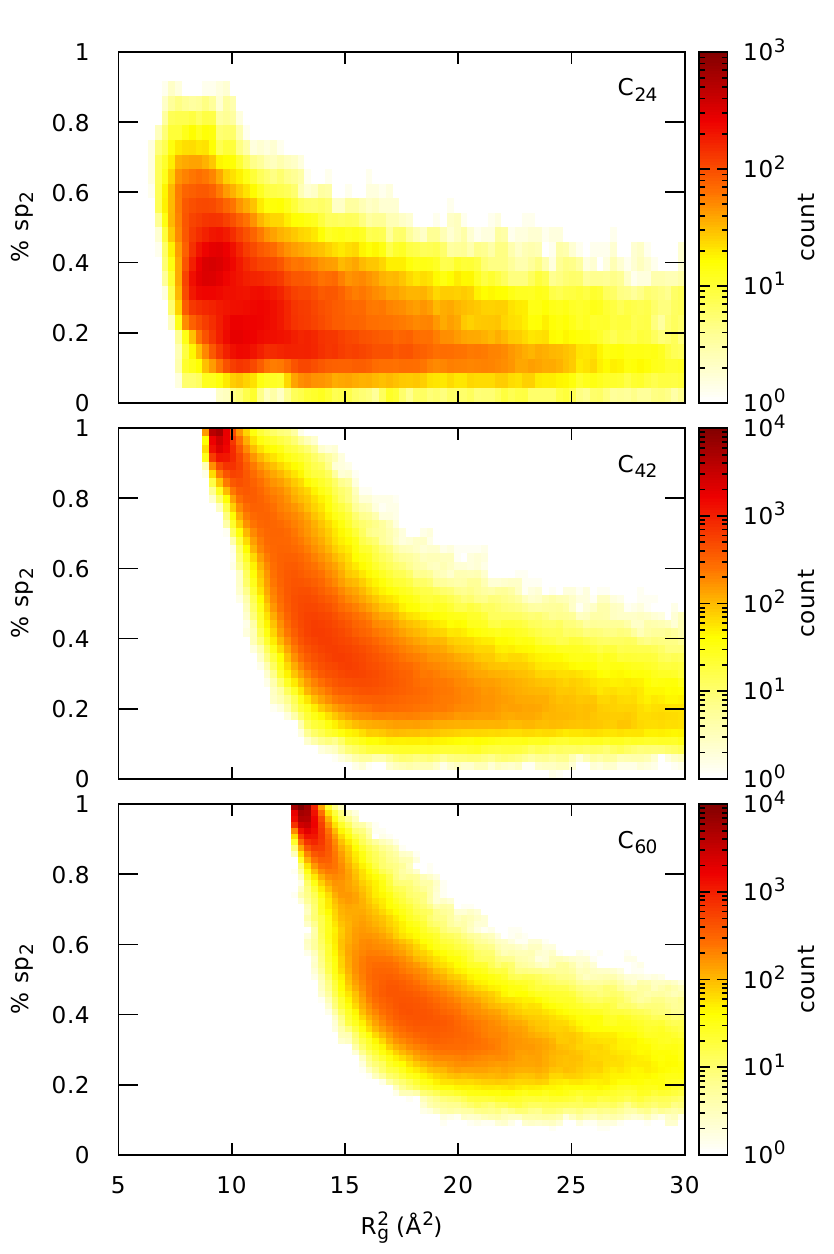}
\caption{(Color online) Distribution of quenched structures for C$_{24}$, C$_{42}$, and C$_{60}$ as a function of the squared gyration radius and the fraction of sp$^2$ carbon atoms.}
\label{fig:dist2}
\end{figure}

\begin{figure}
\centering
\includegraphics[width=8.1cm]{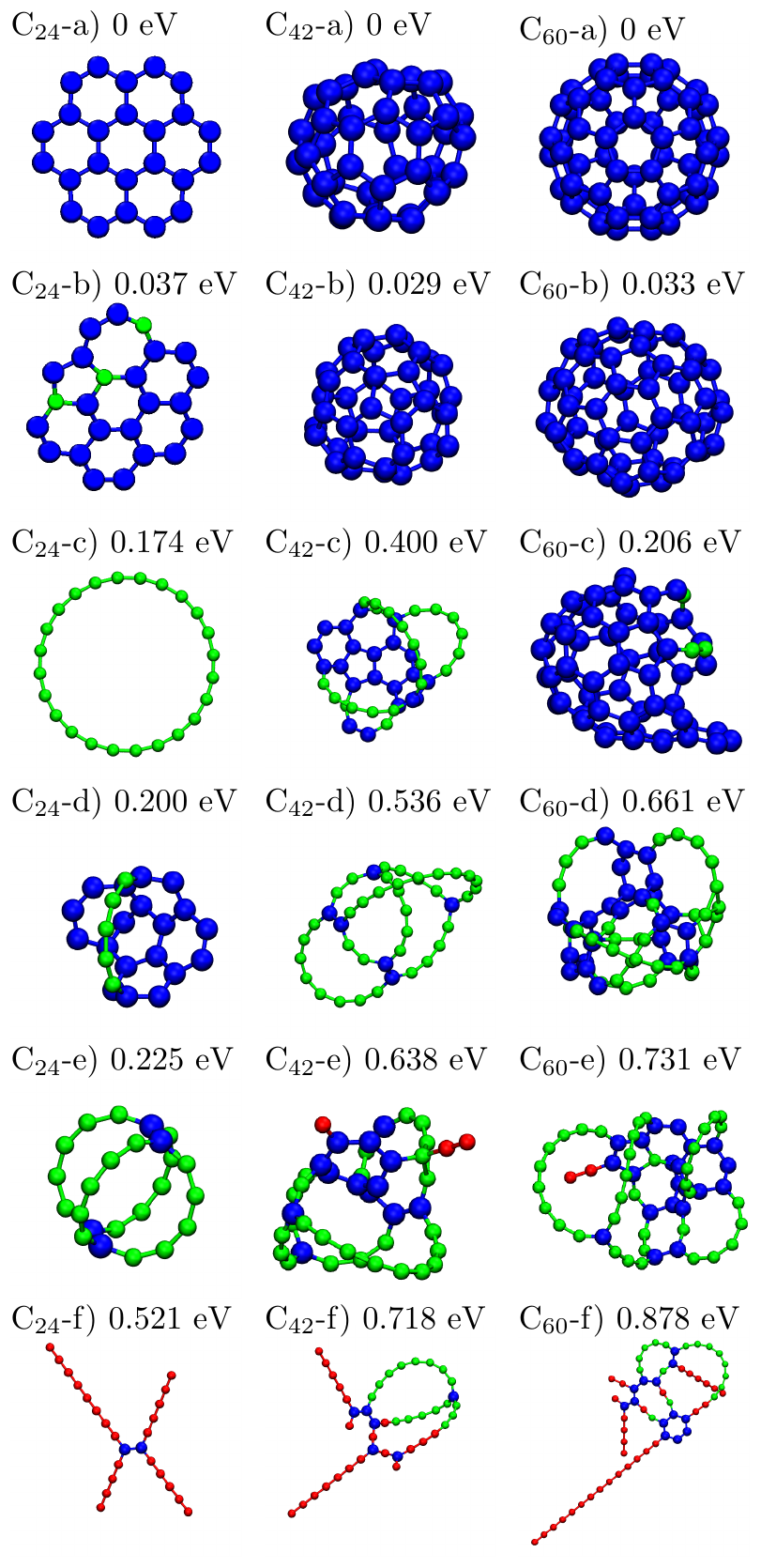}
\caption{(Color online) Selection of quenched structures obtained for C$_{24}$, C$_{42}$, and C$_{60}$. Carbon atoms are colored accordingly with their identified hybridization state, sp and sp$^2$ atoms being shown in red and blue, respectively, and ambiguous atoms in green. No sp$^3$ atom is present in these structures.}
\label{fig:structures}
\end{figure}

The distributions of local minimum energies obtained for the three clusters C$_{24}$, C$_{42}$ and C$_{60}$ with our computational scheme and the REBO potential are represented in Fig.~\ref{fig:distribEp_min}. Here the energies were shifted relative to the reference structure energy to highlight the ranges explored by the REMD simulations, these isomers being depicted at the top of Fig.~\ref{fig:structures} and the ranges themselves highlighted by arrows.
For C$_{24}$, the distribution exhibit a single broad ($\sim$0.2 eV/atom) and asymmetric peak located around 0.31~eV/atom. For C$_{42}$ the distribution exhibits two peaks at 0.11 and 0.56~eV/atom, showing structures on a large energy scale up to $\sim$0.85~eV/atom. Similarly the distribution for C$_{60}$ exhibits two peaks located at 0.11 and 0.77~eV/atom and the overall energies of the structures vary up to $\sim$1~eV/atom. To understand the difference between C$_{24}$ and the two larger cluster sizes and the nature of the two peaks in C$_{42}$ and C$_{60}$, the distributions of quenched structures were determined as a function of the squared radius of gyration and the proportion of sp$^2$ carbon atoms (Fig.~\ref{fig:dist2}).
As for the energy distribution, Fig.~\ref{fig:dist2} shows a single peak for C$_{24}$ and two peaks for C$_{42}$ and C$_{60}$. The lowest energy peaks in C$_{42}$ and C$_{60}$ can be identified to the peak with larger proportion of sp$^2$ and smallest squared gyration radius. The highest energy peaks in C$_{42}$ and C$_{60}$  can be identified to the peak with lowest proportion of sp$^2$ and largest squared gyration radius. A direct examination of the structures allows the lowest-energy peak to be assigned to cage-like structures while the highest-energy peak corresponds to pretzel-like conformations, as already identified by Kim and Tom\'anek \cite{TomanekPRL} who simulated the melting of fullerenes using a tight-binding approach \cite{TomanekPRL}.

In Fig. \ref{fig:structures}, typical structures are shown for different internal energies. At low energy ($\le$0.1~eV/atom), C$_{24}$ tends to be planar as in the reference structure. In this energy range, some deformed caged-structures and various fullerene isomers are mainly found for C$_{42}$ and C$_{60}$. A large proportion of atoms are found as sp$^2$ with only a small number of atoms with ambiguous hybridization. As the internal energy increases, a clear increase in the number of ambiguous hybridization also occurs, concomitant with the appearance of pretzel-like structures around $\sim$0.2 eV/atom for C$_{24}$ and $\sim$0.4 eV/atom for both C$_{42}$ and C$_{60}$.
At higher energies, i.e. $\sim$0.5~eV/atom for C$_{24}$ and $\sim$0.7~eV/atoms for C$_{42}$ and C$_{60}$, a trend to form branched atomic chains with a larger number of sp carbons is seen as well.\par
\begin{figure*}
\centering
\includegraphics{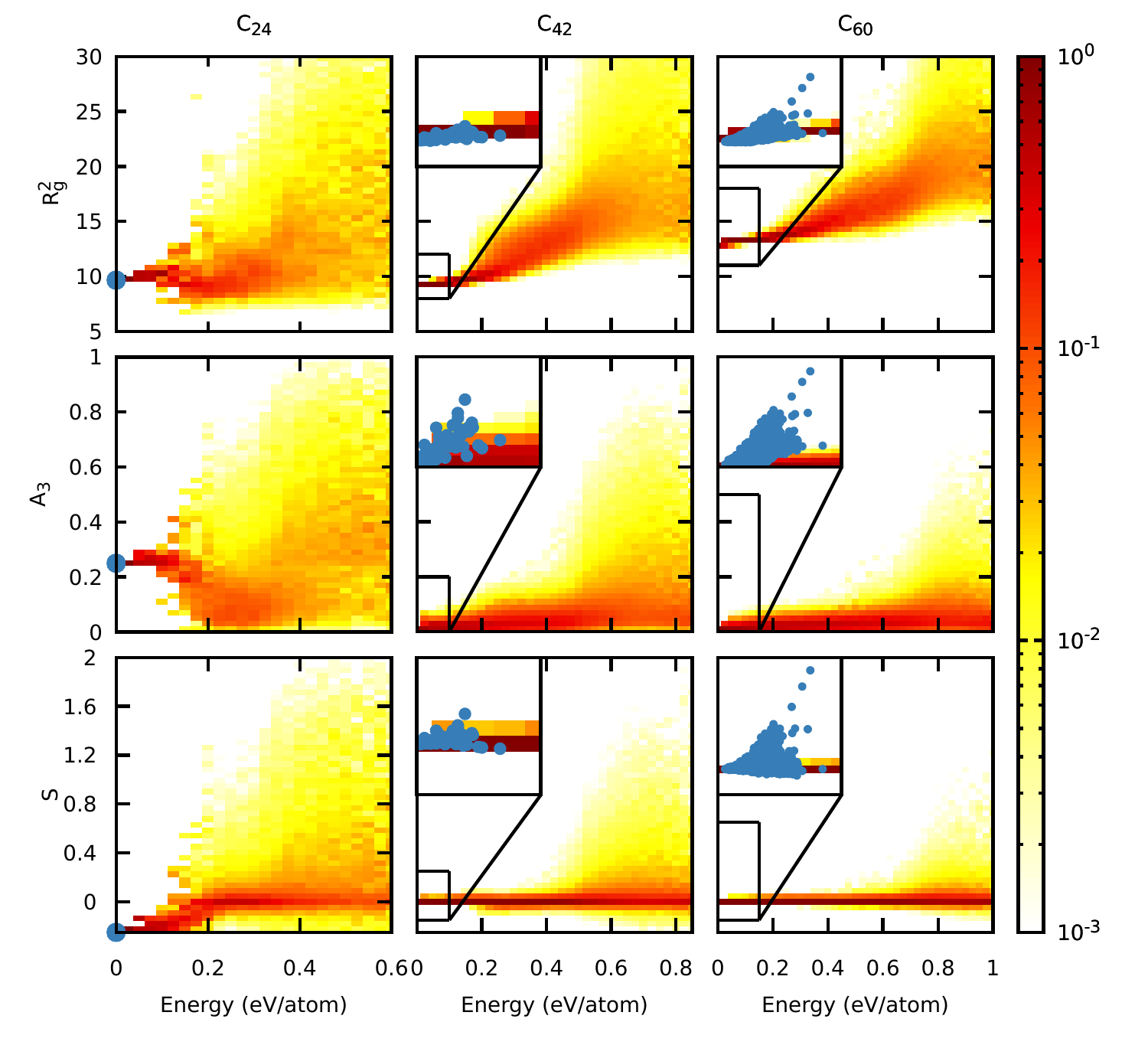}
\caption{(Color online) Distributions of the squared radius of gyration $R_g^2$, asphericity $A_3$ and prolateness $S$ as a function of the energy per carbon atom for C$_{24}$, C$_{42}$, and C$_{60}$. The blue circles correspond to reference structures for C$_{24}$, to the 45 fullerene isomers of C$_{42}$ and to the 1812 fullerene isomers of C$_{60}$.}
\label{fig:gyration}
\end{figure*}

After this first qualitative analysis, we now proceed on a quantitative statistical footing based on the different parameters described in the previous section. The energetic stability of these isomers can be correlated with their structural properties, starting with the global structural order parameters provided by the gyration tensor. The energy-resolved distributions of parameters $R_g^2$, $A_3$ and $S$ are shown in Fig.~\ref{fig:gyration} for the three cluster sizes. At a given energy the distribution is normalized.\par
At low energy, the three structural parameters smoothly converge to the corresponding values in the reference structure. Being shown on the same scale, the square gyration radii are rather similar for C$_{24}$ and C$_{42}$, despite different topologies (planar versus hollow cage). Such differences are better manifested on the asphericity, which vanishes for both cage isomers C$_{42}$ and C$_{60}$. From these 2D distributions the behavior of C$_{24}$ stands out as different from the two larger clusters, with non monotonic variations in the average $R_g^2$ and $A_3$ with increasing isomer energy while the trends are all monotonically increasing in the latter case. A decrease in the square gyration radius near 0.2~eV/atom concomitant with a decrease in $A_3$ indicates spherical structures less extended than fully dehydrogenated coronene and corresponding to the 3D compact structures, as shown in Fig.~\ref{fig:structures}-C$_{24}$-d). Above this energy range, the distribution of squared gyration radius $R_g^2$ becomes much broader and reaches $\sim$20~\AA$^2$.\par
For the two larger clusters, the global structural indices display more regular variations with increasing isomer energy, metastable configurations exhibiting larger gyration radii, a greater asphericity and the prolateness remaining low in magnitude but with increasing fluctuations extending mostly to positive values. These fluctuations are most prominent above 0.5~eV/atom and again convey the greater structural diversity. In contrast, C$_{24}$ remains planar until the energy reaches about 0.2~eV/atom, at which stage the rather sharp variations in the three structural indicators are consistent with the appearance of more compact configurations. In particular, the nearly spherical cage-like structures are found with very low $A_3$ in this energy range.

Above 0.4~eV/atom for C$_{24}$, 0.6~eV/atom for C$_{42}$ and 0.7~eV/atom for C$_{60}$, the asphericity and prolateness indices both explore larger values much deviating from the reference structures. Visual inspection indicates that these high energy isomers are usually branched with several chains and a few rings only, as depicted in Fig.~\ref{fig:structures}-C$_{24}$-f), Fig.~\ref{fig:structures}-C$_{42}$-f),  Fig.~\ref{fig:structures}-C$_{60}$-f). The linear chain isomers, for which $A_3=1$ and $S=2$, were indeed found for C$_{24}$ and for C$_{42}$ but not for C$_{60}$ as they do not fit into the spherical container.

\begin{figure}
\centering
\includegraphics{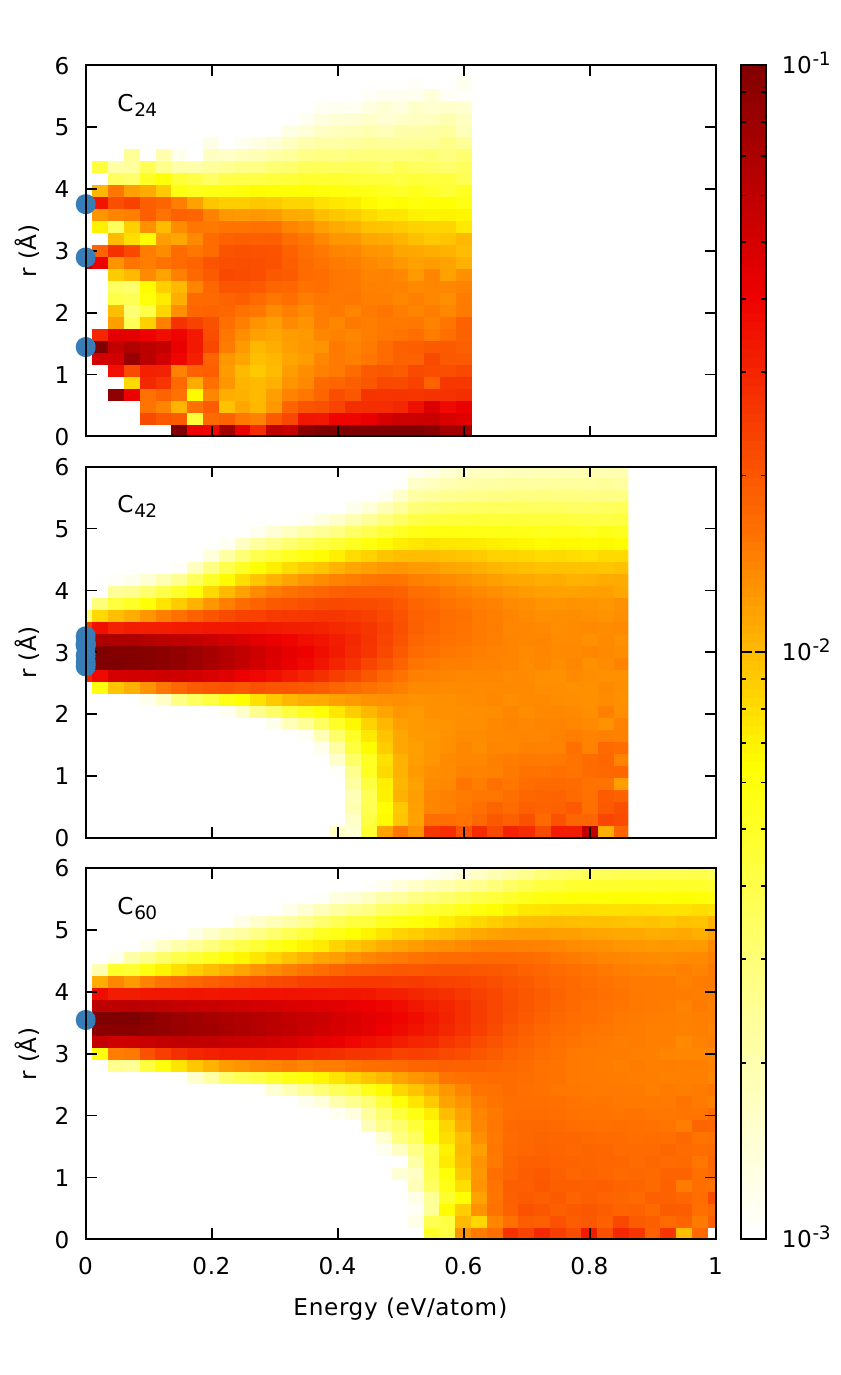}
\caption{(Color online) Radial densities of the structures as a function of the distance from the center of mass, for increasing isomer energy of C$_{24}$, C$_{42}$ and C$_{60}$. The blue circles locate the corresponding peaks in the radial densities of the highly symmetric reference structures.}
\label{fig:radial}
\end{figure}

The radial densities sorted with increasing configuration energy are represented in Fig.~\ref{fig:radial} for the three cluster sizes.
The ranges of variations in the radial density match those exhibited by the global parameters originating from the gyration tensor. In particular, the formation of more compact structures in C$_{24}$ below 0.2~eV/atom, the clear cagelike character in the two larger clusters at low energies manifested by a main peak in the radial density, and the loss of these fullerenic structures above 0.5~eV/atom for C$_{42}$ and 0.6 eV/atom for C$_{60}$ are all reflected in Fig.~\ref{fig:radial}. Interestingly, carbon atoms also much more likely occupy the central regions when the energy exceeds 0.2--0.6~eV/atom depending on system size, which is consistent with the loss of hollow structures and the increasing occurrence of chains and branched configurations.

\begin{figure}
\includegraphics{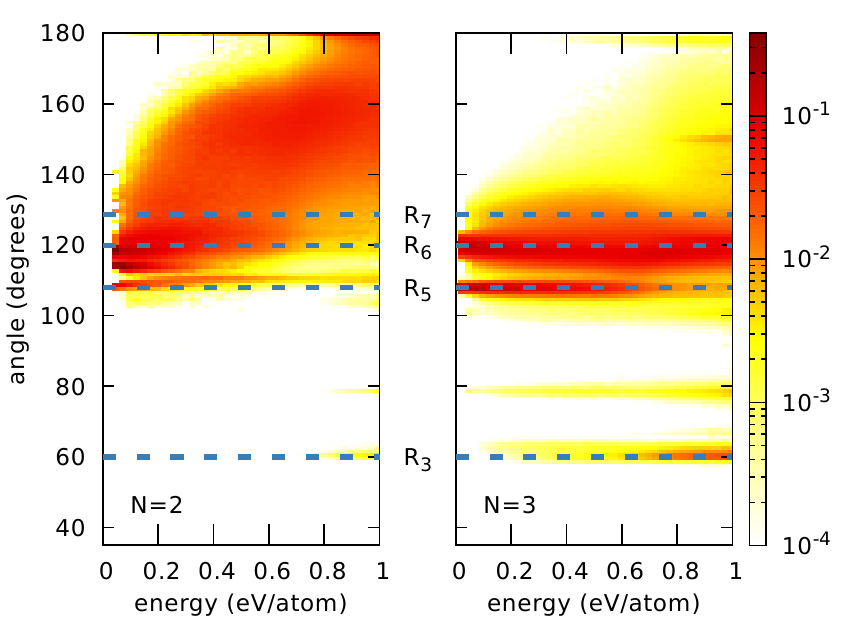}
\caption{(Color online) Angular distribution for carbon atom with two neighbours (left panel) or three neighbours (right panel) in C$_{60}$. The horizontal dashed lines represent the angles for ideal C$_3$ rings (60$^\circ$), R$_5$ rings (108$^\circ$), R$_6$ (120$^\circ$) and R$_7$ rings (128.6$^\circ$).}
\label{fig:angles_c60}
\end{figure}

Turning to hybridization states, we first show in Fig.~\ref{fig:angles_c60} and on the example of C$_{60}$ only how the angles between connected triplets of carbon atoms are distributed when the central atom has two or three neighbors, which we anticipate to be potential candidates as sp and sp$^2$ states, respectively.
The angles corresponding to regular rings R$_3$, R$_5$, R$_6$ and R$_7$ are also shown to highlight the occurrence of such regular polygons in the carbon clusters. At low energies, the angular distribution for carbon atoms having two neighbors only is peaked around 110--120$^\circ$, indicating a majority of pentagonal and hexagonal rings. The angles open at energies higher than 0.3~eV/atom, consistently with the formation or chains or larger rings. Five- and six-membered rings, which are predominant at low energies and in fullerene structures, concomitantly decrease above this same approximate energy threshold. Chains themselves show a signature at an angle of 180$^\circ$. In 3-coordinated atoms, the distribution is strongly peaked around 108$^\circ$ and 120$^\circ$ which are the expected values for buckminsterfullerene composed of perfect pentagonal and hexagonal rings. For both coordination numbers, three-membered rings arise above 0.8~eV/atom, but mostly as traces in 2-coordinated atoms and much more significantly in 3-coordinated atoms as seen through the increasing occurence of 60$^\circ$ angles. This difference suggests that in clusters that are compact enough, the three-membered rings lie in their inner regions. This is corroborated by examining their average distance to the center of mass, which in average approximately equates 83\% of the gyration radius.

\begin{figure}
\centering
\includegraphics{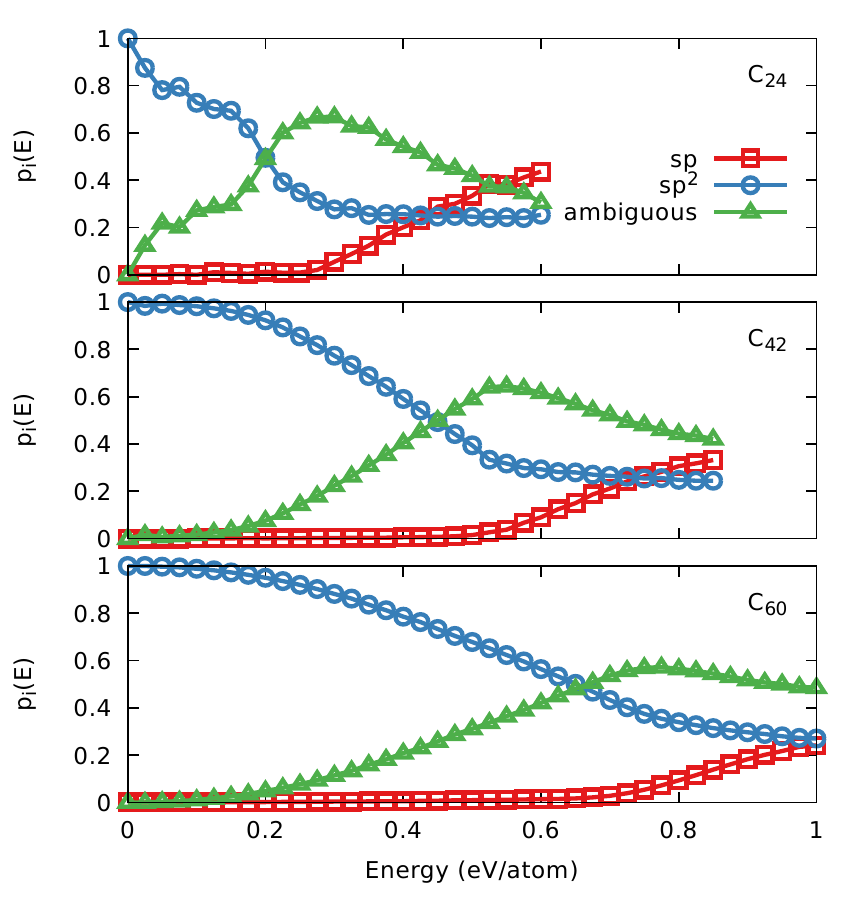}
\caption{(Color online) Probabilities of carbon atoms showing sp, sp$^2$, sp$^3$ or ambiguous hybridization states in the various configurations of C$_{24}$, C$_{42}$ and C$_{60}$, as a function of their energy.}
\label{fig:hybrid}
\end{figure}

Based on the above analysis, the effects of excitation energy on the relative proportion of the various hybridation states can now be discussed. The fractions of sp, sp$^2$, sp$^3$, and ambiguous atoms were thus evaluated for all structures in our databases and for the three cluster sizes, the results being shown in Fig.~\ref{fig:hybrid}.
For the three systems, the reference structures exhibit pure sp$^2$ hybridization state, as expected for the present polyaromatic isomers. As energy increases, the extent of sp$^2$ hybridization drops around energies where the most salient structural changes were noted earlier, that is approximately at 0.2~eV/atom for C$_{24}$, 0.4~eV/atom for C$_{42}$ and 0.6~eV/atom for C$_{60}$. However, sp hybridization becomes significant only at energies higher than these thresholds, while no signature of sp$^3$ is seen whatsoever. Ambiguous hybridization states thus populate the intermediate energy range where configurations become less compact, with a rather high amount of curved linear chains or large rings that do not fall in either of the sp or sp$^2$ categories.

The steady increase in sp hybridization indicates the greater importance of linear chains in high energy structures. At intermediate energies, many configurations show fewer or shorter such chains, at the expense
of rings or curved chains, whose atoms are interpreted as ambiguous until the angle becomes small enough and compatible with sp$^2$ hybridization.

The same qualitative trends are noted in larger clusters, the rise in sp atoms and the maximum in the amount of ambiguous atoms being both shifted to higher energies. Extrapolating these trends, we speculate that in even larger clusters the propensity for sp$^2$ hybridization would become even stronger and more robust against energy excitations, the proportion of linear chains being concomitantly lower and delayed to higher energies.

\begin{figure}
\centering
\includegraphics{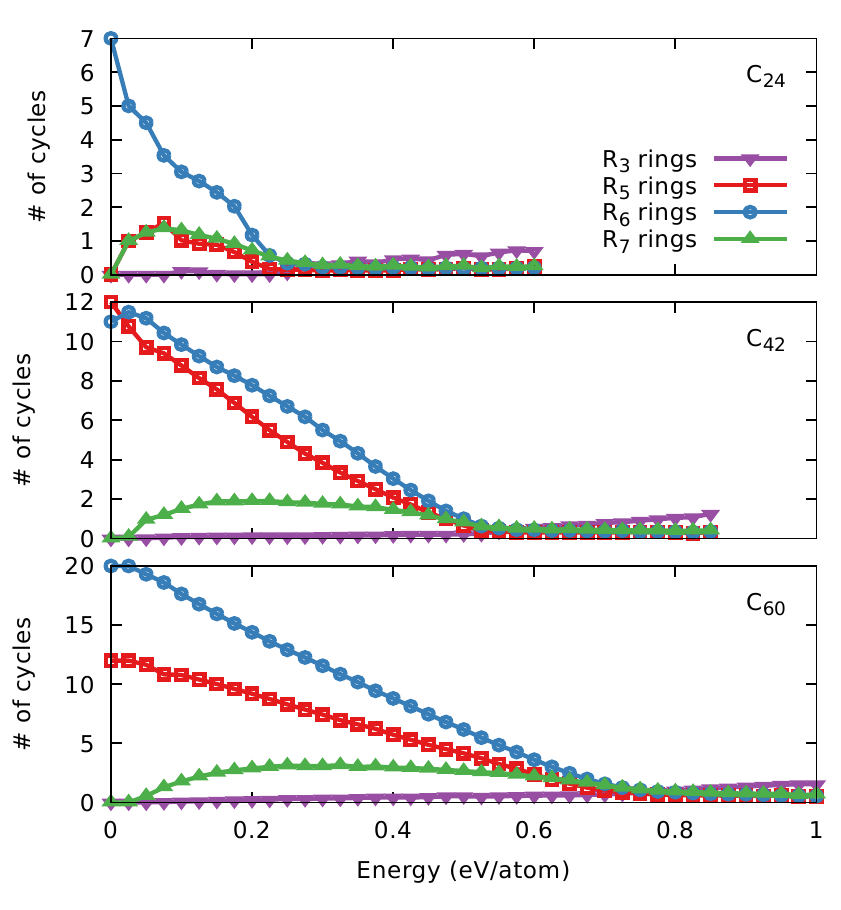}
\caption{(Color online) Average numbers of R$_3$, R$_5$, R$_6$, and R$_7$ rings as a function of isomer energy in C$_{24}$, C$_{42}$, and C$_{60}$. The solid circles at zero energy correspond to the values in the reference structures.}
\label{fig:cycles}
\end{figure}

A complementary quantity is the number and size of the rings contained in the configurations as function of their energy. Figure \ref{fig:cycles} illustrates this specific property, in average, for the 3-, 5-, 6-, and 7-membered rings. 4-membered rings, which are much scarcer, were not considered in this study. Such rings are occasionally found but deviate significantly from the perfect square and exhibit angles closer to 80$^\circ$ (and a more standard angle near 120$^\circ$). In the case of C$_{60}$ they can be seen in Fig.~\ref{fig:angles_c60} for three-coordinated atoms as an horizontal spot with very low magnitude near 80$^\circ$.
For the three clusters, the reference structures only contain 5- and 6-membered rings. In all cases, the number of hexagonal rings steadily decreases with increasing energy. In both C$_{24}$ and C$_{60}$ this decrease benefits to pentagonal and heptagonal rings, consistently with the appearance of topological defects such as Stone-Wales pairs.

As the internal energy reaches the threshold values where global structural changes take place (0.2~eV/atom for C$_{24}$, 0.5~eV/atom for C$_{42}$, 0.7~eV/atom for C$_{60}$), the numbers of rings having 5 or more atoms reaches a minimum and only residual 3-rings are found although with an increasing propensity. The loss of large rings is consistent with more atoms being present as linear chains in such structures, as depicted in Fig.~\ref{fig:structures}. Three-membered rings being energetically rather disfavored, they only appear --- occasionally --- at the highest excitation energy as a way to connect linear chains into branched structures.

\begin{figure}
\centering
\includegraphics{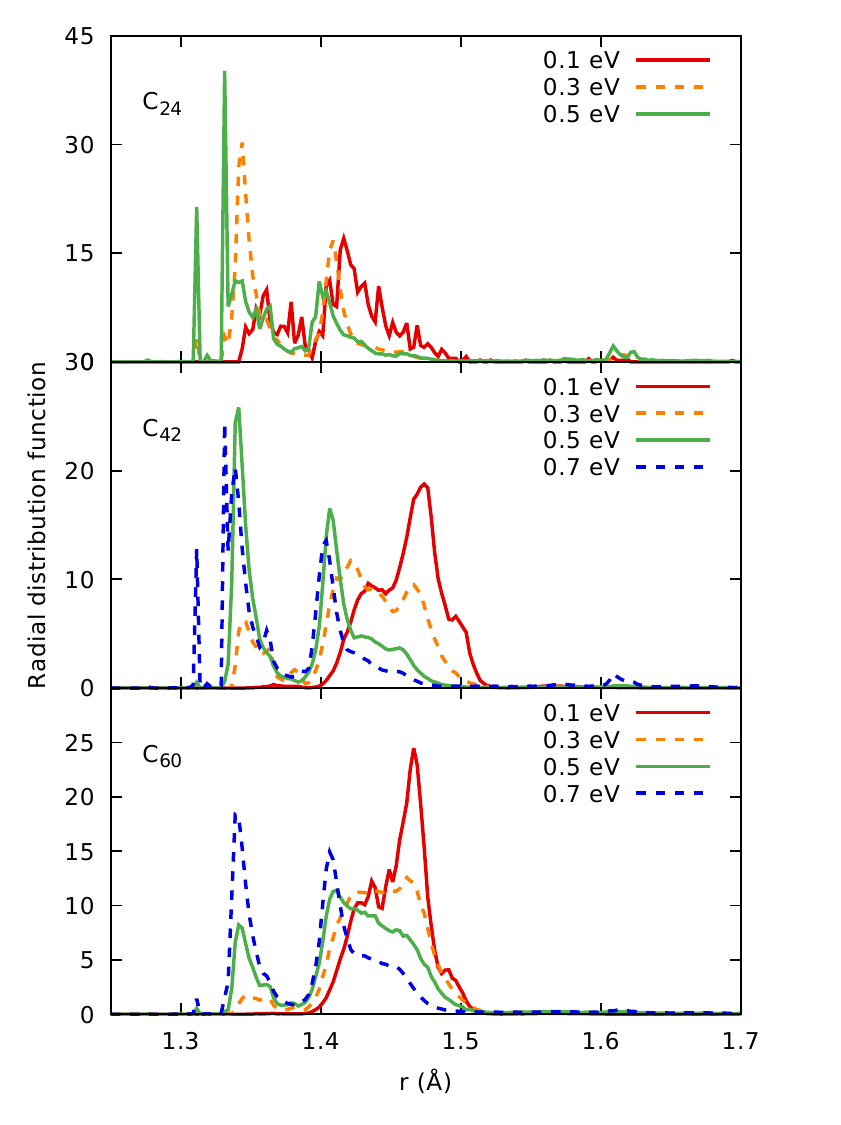}
\caption{(Color online) Pair distribution functions as a function of the carbon-carbon distance for C$_{24}$, C$_{42}$, and C$_{60}$ at various energies.}
\label{fig:rdf}
\end{figure}

We have finally considered the pair distribution function as another structural indicator sensitive to the chemical arrangement and in particular hybridization state. Such quantities are shown in Fig.~\ref{fig:rdf} for the three cluster sizes and at selected energies.
At low energy of 0.1~eV/atom and for all sizes, the pair distribution function is dominated by sp$^2$ carbons. For C$_{42}$ and C$_{60}$, in the range 1.4--1.5~\AA\, carbon atoms are  involved in hexagonal and heptagonal rings with the two corresponding peaks located at 1.43 and 1.47~\AA, respectively. Note that for buckminsterfullerene the CC distance for a bond located between two hexagonal ring is 1.40~\AA\, while bonds connecting  hexagonal and pentagonal rings together are closer to 1.45~\AA. For the C$_{42}$ reference structure with $D_3$ point group, the CC distance between two hexagonal rings is 1.41~\AA, the CC distance between an hexagonal and a pentagonal ring varies between 1.45 and 1.46~\AA. Finally the CC distance between two pentagonal rings varies between 1.48 and 1.49~\AA. For C$_{24}$, the CC distance of the $D_{6h}$ reference structure for the inner bonds varies between 1.44~\AA\ and 1.45~\AA. For the outer bond the distance varies between 1.37~\AA\ and 1.41~\AA.

As energy increases, ambiguous and sp hybridization states become increasingly important and are manifested by a narrower peak near 1.34~\AA, and eventually 1.31~\AA\ for singly coordinated atoms. A small residual peak above 1.6~\AA\ is also found at very high energy. This peak results from the few carbon atoms involved in three-carbon rings.

The fullerene cage C$_{60}$ and its cation were both recently observed in the ISM. Identification of the neutral buckminsterfullerene could be achieved owing to its few and very specific active bands \cite{Omont2016}. The observation of other clusters, even of the fullerene form, should be far more challenging. One important issue is to determine how such highly organized molecules could be formed in extremely dilute media, notably in the top-down assumption where C$_{60}$ would originate from dehydrogenation and subsequent rearrangement of larger polycyclic aromatic hydrocarbons. Under such a model, our results indicate that the route to buckminsterfullerene is far from straightforward and encompasses multiple branched isomers along the way. It is noteworthy that the high energy configurations found in this work resemble the pretzel phase previously identified by Kim and Tom\'anek \cite{TomanekPRL} in their simulation of fullerene melting. With the threshold energy reported here, the pretzel phase of C$_{60}$ might be present as well in the ISM.

\section{Statistical data analysis}

In the previous section the relation between isomer energy and various structural parameters was highlighted, emphasizing the large configurational diversity arising as energy increases. Here we adopt a different but complementary point of view by questioning to which extent  energetic stability is statistically related to these geometric parameters. More precisely, we have performed a systematic correlation analysis between isomer energy and the square gyration radius $R_g^2$, the asphericity $A_3$, the prolateness $S$ and the sp$^2$ fraction, for the three databases of minima obtained for C$_{24}$, C$_{42}$, and C$_{60}$.

The most straightforward way to quantify {\em linear} correlation between two sets of data is based on the traditional Pearson's coefficient that we denote $R_{ij}$ for two sets of variables $X_i$ and $X_j$ among the five
quantities of isomer energy, $R_g^2$, $A_3$, $S$ and sp$^2$ fraction. We are mostly interested in correlations between energy (acting as $X_i$, our output parameter) and any of the four other quantities (acting as $X_j$ for $j\neq i$ and treated as input parameters).

One known issue with Pearson correlation coefficients is that they incorporate possibly strong correlations between the various input parameters, and an efficient way of removing these correlations consists of considering partial correlation coefficients (PCCs) \cite{hamby94} instead. The PCC between variables $X_i$ and $X_j$ involves the full linear correlation matrix ${\bf R}=R_{ij}$ (including also those elements $R_{jk}$ between geometrical parameters $X_j$ and $X_k$ for $k\neq i$) and reads
\begin{equation}
\mbox{PCC}(\{X_i\},\{X_j\}) = -\frac{P_{ij}}{\sqrt{P_{ii} P_{jj}}},
\label{eq:pcc}
\end{equation}
where $P_{ij}$ is the cofactor of the element $R_{ij}$ in the determinant $|{\bf R}|$ of matrix $\mathbf{R}$.

The linear ($R_{ij}$) and partial correlation coefficients obtained between the four geometrical parameters and the isomer energy are given in Table~\ref{table:pcc} for the three cluster sizes.
\begin{table}
\begin{tabular*}{\linewidth}{@{\extracolsep{\fill}} crrrrrr}
\hline
\hline
& \multicolumn{2}{c}{C$_{24}$}
& \multicolumn{2}{c}{C$_{42}$}
& \multicolumn{2}{c}{C$_{60}$} \\
\cline{2-3}\cline{4-5}\cline{6-7}
& $R$ & PCC & $R$ & PCC & $R$ & PCC \\
\hline
$R_g^2$ & 0.436 & -0.217 & 0.517 & 0.078 & 0.623 & 0.183 \\
$A_3$ & 0.365 & 0.056 & 0.444 & 0.112 & 0.433 & 0.103 \\
$S$ & 0.313 & 0.085 & 0.272 & 0.066 & 0.206 & 0.031 \\
sp$^2$ & -0.285 & -0.114 & -0.906 & -0.852 & -0.965 & -0.933 \\
\hline
\hline
\end{tabular*}
\caption{\label{table:pcc} Linear ($R$) and partial correlation coefficients (PCC) between various parameters and isomer energy obtained from the databases of minima for C$_{24}$, C$_{42}$ and C$_{60}$.}
\end{table}
From these data, the smaller cluster C$_{24}$ appears to behave somewhat differently from the two larger clusters C$_{42}$ and C$_{60}$. For these sizes that support fullerene-type isomers, linear correlations seen from $R$ quantities are highest with the sp$^2$ fraction, the negative sign obtained for $R$ being consistent with the intuitive observation that isomers lowest in energy have the highest sp$^2$ fraction. The three shape parameters always correlate positively but with Pearson coefficients never exceeding 0.65, such a poor linear correlation appears consistent with the scatter plots reported in Fig.~\ref{fig:gyration}.

Removing the correlations between the various parameters in the PCC quantities confirms this trend for the two larger clusters, and highlights the sp$^2$ fraction as the most sensitive parameter causing relative energetic stability among conformers. However, for C$_{24}$ this quantity performs not as well, and none of the quantities shows a strong partial correlation with energy. This can be understood because for C$_{24}$, within the main peak of the energy distribution (between 0.2 and 0.6 eV/atom) the sp$^2$ fraction remains relatively constant. The negative sign obtained with the square gyration radius is related with the markedly different distributions displayed in Fig.~\ref{fig:gyration}, where for C$_{24}$ structures near 0.2~eV/atom are found to be more compact than the planar global minimum, at variance with the behavior noted for the two larger clusters.

The present statistical analysis thus suggests that, for fullerenes, the fraction of sp$^2$ atoms is the main factor responsible for the energetic stability of the various conformers, geometric shape parameters appearing of lesser importance. Proceeding further, we have attempted to represent the energy $X_0$ as a simple function of the other parameters $\{X_i\}$, $1\leq i\leq 4$, using a simple linear functional form as
\begin{eqnarray}
    \widetilde E &=& a_0 + \sum_i a_i X_i + \sum_{i<j} a_{ij} X_iX_j \nonumber \\
    &&+\sum_{i<j<k} a_{ijk} X_i X_j X_k + a_{1234} X_1 X_2 X_3 X_4,
    \label{eq:efit}
\end{eqnarray}
where the 16 parameters $a$ are obtained by minimizing exactly the least square penalty function $|\widetilde E-X_0|^2$ summed over all minima.

\begin{figure}
    \centering
    \includegraphics[width=6.5cm]{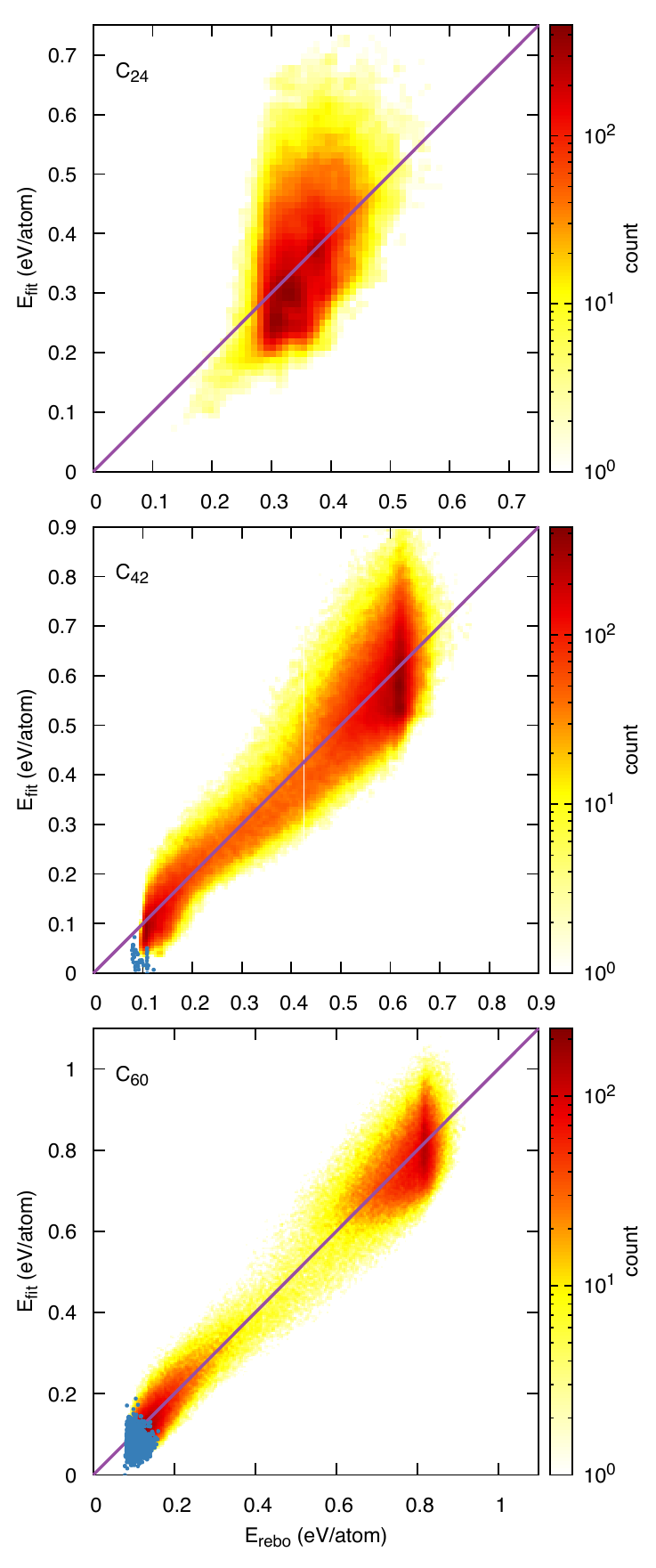}
    \caption{(Color online) Correlations between fitted energy as a function of the REBO energy (eV/atom) for the databases of minima of C$_{24}$ (upper panel), C$_{42}$ (middle panel), and C$_{60}$ (lower panel). For C$_{42}$ and C$_{60}$ the smaller sets of fullerene cages is superimposed as scatter plots with blue symbols.}
    \label{fig:correE}
\end{figure}

The optimized fitted form for the isomer energy is shown in Fig.~\ref{fig:correE} against the reference value, for the three systems and as a 2D distribution.
These plots generally reflect the shapes of the distributions in Fig.~\ref{fig:distribEp_min}, which are bimodal for the two larger clusters but only unimodal for C$_{24}$. Even after optimizing the parameters, the representation of the isomer energy as a function of the three shape indicators $R_g^2$, $A_3$, $S$ together with the sp$^2$ fraction appears rather approximate, especially for the fullerenes subsets. However, the trends are satisfactory with Pearson coefficients between the REBO and the fitted energies exceeding 0.95 for both C$_{42}$ and C$_{60}$. The smaller cluster C$_{24}$ actually exhibits less chemical ordering and weaker correlations between energy and all other quantities analyzed, hence our representation of the energy as a linear form of these parameter unsurprisingly performs poorly.

\section{Concluding remarks}

Owing to their additional free surface, carbon clusters are anticipated to exhibit a structural diversity that matches or even exceeds that of the known allotropy of bulk carbon. The present article was aimed at quantifying this diversity from a computational perspective, by performing a broad unbiased sampling of low-energy structures of carbon clusters containing a few tens of atoms. For this purpose the reactive semi-empirical REBO potential was employed for its ability to describe all hybridization states of carbon atoms together with REMD simulations as our main exploration tool. The simulations were further processed by systematically performing local optimizations in order to uncover the inherent structures and sort them as a function of energy. Various structural indicators of global or local character were evaluated to relate the structural and energetic features of these configurations to one another.

The lowest-energy structures obtained for the three sizes of interest here with 24, 42, or 60 atoms were found to be mostly polyaromatic and either planar as in the case of the fully dehydrogenated coronene flake for C$_{24}$, or cage-like as in C$_{42}$, buckminsterfullerene providing the most symmetric case of a spherical structure for C$_{60}$. Excess energy appears as the main driving force causing the structural diversity of other isomers, and we evaluated 0.2~eV/atom as a threshold energy above which this diversity explodes, leading to a mixture of isomers that are neither fully planar nor perfectly cage-like but contain an increasing number of rings and connecting chains. Along this transformation the atoms initially keep a mostly sp$^2$ character but evolve toward a greater proportion of sp hybridization (through an intermediate state where hybridization is ambiguous), no significant sp$^3$ character being noted. It should be noticed that the spherical container used here to restrict to prevent excessively dissociated structures also  disfavors elongated but connected chains as well as large planar structures for the bigger clusters, which could be entropically favored at high energies and thus be of significance as reaction intermediates. Periodic boundary conditions at fixed density or pressure or a Monte Carlo framework restricted to sample connected configurations only could complement the present approach and possibly generate new relevant configurations.

A systematic statistical analysis of the databases of minima gathered for the three systems was performed using linear and partial correlation coefficients in order to unravel a possible sensitivity of the relative isomer energy to the various indicators evaluated. For the two clusters supporting fullerenes, the sp$^2$ fraction was found as the main driving force correlating with isomer stability, while none of the shape parameters was found particularly relevant. A linear representation of the energy as a function of the remaining parameters was proposed as an approximate description.

In view of astrophysical implications, one extension of the present work could consider the effects of a single charge on the structure of such carbon clusters. Ionized clusters are much more convenient to study in laboratory experiments owing to mass spectrometry selection, and their structures can be indirectly measured by techniques such as ion mobility \cite{Helden1991,Helden1993}. Cationic clusters could be modeled either by incorporating appropriate electrostatic corrections to the present REBO model, e.g. through fluctuating charges \cite{goddard}, or by approaches explicitly accounting for electronic structure such as tight-binding \cite{Van-Oanh:2002oq,TomanekPRL} or DFTB \cite{dftb}.

Besides their ionic character, the pure carbon clusters studied here are an oversimplified description of the chemical ISM. The presence of hydrogen should be considered, especially in relation with the presence of polycyclic aromatic hydrocarbons as the starting point toward fullerene formation. To cope with the additional chemical complexity resulting from the presence of unlike atoms, improvements of the present REMD method or alternative approaches such as basin-sampling could be beneficial. Such investigations about the possible contribution of hydrogen at selected fractions will be the addressed in the near future.

\section{Acknowledgments}
The authors gratefully acknowledge financial support by the Agence Nationale de la Recherche (ANR) grant ANR-16-CE29-0025. 
This work was supported by grants from R\'egion Ile-de-France and by the GDR EMIE 3533.

\bibliography{biblio}

\end{document}